\documentclass{aa}  
\usepackage{graphicx}
\usepackage{txfonts}

\usepackage{caption}
\usepackage{multicol}      
\usepackage{multirow}
\usepackage{natbib}
\bibpunct{(}{)}{;}{a}{}{,} 

\usepackage[colorlinks=true, urlcolor=blue, colorlinks=true, citecolor=blue, linkcolor=blue]{hyperref}

\begin{document} 

  \title{Jeans criterion in hydrostatic stratified media}
   
   \author{Mahmood Roshan
          \inst{1}\thanks{\email{mroshan@um.ac.ir}} \and Asiyeh Habibi \inst{1} 
          }

   \institute{Department of Physics, Faculty of Science, Ferdowsi University of Mashhad, P.O. Box 1436, Mashhad, Iran 
             }

   \date{\today}
   
\abstract
{}
{The classical Jeans stability criterion neglects spatial gradients in background physical quantities such as density and pressure. Here, we revisit the Jeans analysis for a nonrotating fluid in hydrostatic equilibrium, explicitly accounting for these gradients and deriving a modified dispersion relation governing perturbation propagation.}
{Using physical arguments, we show why the Jeans swindle is not required to derive the Jeans criterion for perturbations that are small compared to the size of the host system. We use linear perturbation analysis alongside an averaging procedure to study the behavior of the local Eulerian perturbations in a hydrostatic medium. We show that background gradients enter the dispersion relation in a highly nontrivial manner, precluding the derivation of a general gravitational stability criterion applicable to an arbitrary hydrostatic system. }
{We demonstrate that, in the local short-wavelength limit, the standard Jeans criterion remains valid despite the presence of nonzero background gradients. Furthermore, we show that all local perturbations in a hydrostatic system are stable against local gravitational collapse.}
{We conclude that it is not possible to derive a general local gravitational stability criterion valid for an arbitrary hydrostatic background. However, in the local short-wavelength limit, the standard Jeans criterion remains valid.}


   \maketitle

\section{Introduction}\label{introduction}

The local gravitational stability of a self-gravitating system was first studied by Sir James Jeans in 1902 \citep{jeans}. In that work, the density and pressure of the infinite background medium were assumed to be constant. It is well known that such a homogeneous and infinite self-gravitating system does not exist in reality. Despite this unphysical assumption, the final Jeans stability criterion has proven extremely useful and widely used in astrophysical studies. However, the inconsistent background solution used in the analysis, leading to the so-called Jeans swindle \citep{binney}, has long remained challenging. Several studies have attempted to justify the swindle and explain the remarkable practical validity of the Jeans stability criterion \citep{Bonnor1957,s2,s3,s4}. 

Before introducing a new dispersion relation for gravitational stability, we first demonstrate, on physical grounds, that no Jeans-swindle assumption is required and that the classical Jeans criterion remains valid for perturbations whose characteristic scales are sufficiently small compared with the characteristic scale of the host system for the background gravitational field to have a negligible effect on their internal dynamics. The novel contribution of this paper then consists in deriving a revised dispersion relation governing the gravitational stability of astrophysical systems in hydrostatic equilibrium, explicitly accounting for non-negligible density and pressure gradients as well as the non-negligible background self-gravity. In many astrophysical environments, such as stellar interiors, molecular cloud cores, and stratified gaseous systems, density and pressure gradients are inherently non‑negligible, which motivates a reassessment of the classical Jeans analysis \citep{larson,ostriker}. For recent studies on the stability criterion in stratified media, we refer the reader to \cite{2023MNRAS.518.5154N,nipoti}.

The outline of this paper is as follows. In Section~\ref{section1}, we show, using physical arguments, that no swindle is required in the Jeans analysis and that the standard Jeans stability criterion holds when the perturbation scale is small compared to the characteristic length of the host system. In Section~\ref{section2}, we relax this assumption by accounting for the spatial gradients of the background quantities, leading to a new dispersion relation, which we study in the local short-wavelength limit to derive the gravitational stability criterion. The conclusions are presented in Section~\ref{section4}.

\section{No need for the Jeans swindle}\label{section1}
We begin with the equations that govern the evolution of perturbations in a self-gravitating fluid system:
\begin{equation}
\frac{\partial \mathbf{v}}{\partial t}+(\mathbf{v}\cdot\nabla)\mathbf{v}=-\frac{\nabla p}{\rho}-\nabla \Phi,
\end{equation}
\begin{equation}
\frac{\partial \rho}{\partial t}+\nabla\cdot (\rho \mathbf{v})=0,
\end{equation}
\begin{equation}
\nabla^2\Phi=4\pi G \rho.
\end{equation}
The first equation is Euler's equation, the second is the continuity equation, and the last one is Poisson's equation. $\mathbf{v}$, $p$, $\rho$, and $\Phi$ denote the velocity, pressure, mass density, and gravitational potential, respectively. Throughout this study, we assumed a barotropic equation of state, namely $p=p(\rho)$.  We considered a hydrostatic background characterized by density $\rho_0$ and pressure $p_0$, satisfying $\nabla p_0/\rho_0=-\nabla \Phi_0$. We then perturbed the background quantities as $p=p_0+p_1$, $\rho=\rho_0+\rho_1$, $\mathbf{v}=\mathbf{v}_1$ and $\Phi=\Phi_0+ \Phi_1$, where the subscript ``1'' denotes linear Eulerian perturbations. The first-order perturbation equations were then obtained::
\begin{equation}\label{l1}
\frac{\partial \mathbf{v}_1}{\partial t}=\frac{\nabla p_0}{\rho_0}\frac{\rho_1}{\rho_0}-\frac{\nabla p_1}{\rho_0}-\nabla \Phi_1,
\end{equation}
\begin{equation}\label{l2}
\frac{\partial \rho_1}{\partial t}+\nabla\cdot (\rho_0 \mathbf{v}_1)=0,
\end{equation}
\begin{equation}\label{l3}
\nabla^2\Phi_1=4\pi G \rho_1,
\end{equation} 
In the standard Jeans analysis, it is customary to assume that $p_0$ and $\rho_0$ are constant. By differentiating equation \eqref{l2} with respect to time and making use of equations \eqref{l1} and \eqref{l3}, one readily finds:
\begin{equation}\label{l4}
\frac{\partial^2 \rho_1}{\partial t^2}-c_s^2 \nabla^2 \rho_1-4\pi G \rho_0 \rho_1=0,
\end{equation}
where $c_s$ is the sound speed in the system, which remains constant throughout due to the conditions imposed in the analysis. The Fourier transform of the density perturbation is given by $\rho_1(\mathbf{r},t)=\frac{1}{(2\pi)^4}\int \hat{\rho}_1(\omega',\mathbf{k'})e^{i(\mathbf{k'}\cdot\mathbf{r}-\omega' t)}d^3k'd\omega'$. Multiplying equation \eqref{l4} by $e^{i(\mathbf{k}\cdot\mathbf{r}-\omega t)}$ and integrating over all space and time $t$ yields the dispersion relation $\omega^2=c_s^2 k^2-4\pi G\rho_0$. This directly leads to the stability condition $k\geq k_{\rm J}$, where $k_{\rm J}=\sqrt{4\pi G\rho_0/c_s^2}$ is the standard Jeans wavenumber. In other words, perturbations with wavenumbers smaller than $k_{\rm J}$ are unstable. However, the difficulty with this analysis is that the background configuration with $\mathbf{v}_0=0$, $p_0=\mathrm{const.}$, and $\rho_0=\mathrm{const.}$ is inconsistent: the Euler equation implies $\nabla\Phi_0=0$, while Poisson’s equation requires $\nabla\cdot\nabla\Phi_0\neq0$. This inconsistency has led to the suggestion that there is a swindle in the analysis. In the following, we show that no such swindle is required to derive the stability criterion.

We compared the first and second terms on the right-hand side (RHS) of equation \eqref{l1}. To do so, it was necessary to compare $\nabla p_0$ with $\nabla p_1$. When considering a small perturbation relative to the size of the host system, the pressure perturbation $p_1$ varies rapidly over a short length scale comparable to the characteristic size of the perturbation, whereas the background pressure does not exhibit such rapid spatial variations within that same region. In other words, it is reasonable to expect that $|\nabla p_0| \ll |\nabla p_1|$. Therefore, noting that the coefficient $|\rho_1/\rho_0|\ll1$, we conclude that the first term on the RHS of \eqref{l1} can be neglected in comparison with the second term. 

We can reach the same conclusion by comparing the first and third terms on the RHS of \eqref{l1}. To do so, we employed the hydrostatic condition and replaced $\nabla p_0/\rho_0$ with $-\nabla\Phi_0$. Consequently, we then needed to compare the background gravitational field $\nabla\Phi_0$ with that of the perturbation $\nabla \Phi_1$. When the perturbation size is small relative to the overall system, the background gravitational field does not significantly influence the internal dynamics of the perturbation, just as the gravitational field of the Milky Way does not significantly affect the motion of the planets within the solar system around the Sun. It only determines the motion of the perturbation’s center of mass. From this perspective, the first term on the RHS of \eqref{l1} is negligible compared to the last term, namely $|\nabla \Phi_0| \ll |\nabla \Phi_1|$.

Considering the assumption above regarding the characteristic size of the perturbation, one may infer that the gradients of both the background density and the sound speed can also be neglected across the perturbation. Consequently, we again arrive at equation \eqref{l4} and recover the same Jeans stability criterion. The distinction, however, is that in this case the derivation does not require invoking the Jeans swindle.

\section{Jeans stability criterion revisited}\label{section2}
In this section, we relax the main assumption underlying the standard Jeans analysis. In other words, we no longer neglect the local gradients of the background quantities. In this case, one can show that equation \eqref{l4} is modified as follows:
\begin{equation}\label{l5}
\begin{split}
\frac{\partial^2 \rho_1}{\partial t^2}-& \nabla^2 (c_s^2 \rho_1)-4\pi G \rho_0 \rho_1\\&+\nabla\cdot\big(\frac{\nabla p_0}{\rho_0} \rho_1\big)-\nabla\Phi_1\cdot\nabla\rho_0=0,
\end{split}
\end{equation}
which recovers the standard case when $\nabla p_0=0$, $\nabla\rho_0=0$ and $\nabla c_s^2=0$.

The Fourier transforms of the background functions and perturbations are given by
\begin{equation}\label{j11}
Q(\mathbf{r})=\frac{1}{(2\pi)^3}\int \hat{Q}(\mathbf{k'})e^{i\mathbf{k'}\cdot\mathbf{r}}d^3 k',
\end{equation}
\begin{equation}\label{j12}
Q_1(\mathbf{r},t)=\frac{1}{(2\pi)^4}\int \hat{Q}_1(\omega',\mathbf{k'})e^{i(\mathbf{k'}\cdot\mathbf{r}-\omega' t)}d^3 k' d\omega',
\end{equation}
where $Q$ denotes $\rho_0$, $c_s^2$, or $\Phi_0$, while $Q_1$ denotes $\rho_1$ or $\Phi_1$. Using Poisson's equation, we obtained
\begin{equation}\label{j13}
\hat{\Phi}_0(\mathbf{k})=-\frac{4\pi G}{k^2} \hat{\rho}_0(\mathbf{k}),
\end{equation}
\begin{equation}\label{l6}
\hat{\Phi}_1(\omega,\mathbf{k})=-\frac{4\pi G}{k^2} \hat{\rho}_1(\omega,\mathbf{k}).
\end{equation}
We then substituted the Fourier transforms given by equations \eqref{j11} and \eqref{j12} into equation \eqref{l5}. The resulting expression was then multiplied by $e^{-i(\mathbf{k}\cdot\mathbf{r}-\omega t)}$ and integrated over all space and time. In the following, we illustrate the procedure by evaluating three terms in \eqref{l5}. For the first term in \eqref{l5}, we obtain
\begin{equation}
\int \frac{\partial^2 \rho_1}{\partial t^2} e^{-i(\mathbf{k}\cdot\mathbf{r}-\omega t)} d^3x\, dt=-\omega^2 \hat{\rho}_1(\omega,\mathbf{k}),
\end{equation}
where we used
\begin{equation}\label{newd1}
\begin{split}
&\int e^{i(\mathbf{k}-\mathbf{k}')\cdot \mathbf{r}}d^3 x=(2\pi)^3\delta^3(\mathbf{k}-\mathbf{k}'),\\&
\int e^{i(\omega-\omega')t}dt=2\pi \delta(\omega-\omega').
\end{split}
\end{equation}
We then considered the term involving $\rho_0\rho_1$. In this case, we obtained
\begin{equation}
\begin{split}
&\int \rho_0 \rho_1 e^{-i(\mathbf{k}\cdot\mathbf{r}-\omega t)} d^3x\, dt=\frac{1}{(2\pi)^7} \times \\& \int \hat{\rho}_0(\mathbf{k}')\hat{\rho}_1(\omega'',\mathbf{k}'') e^{i(\mathbf{k}''+\mathbf{k'}-\mathbf{k})\cdot \mathbf{r}} e^{i(\omega-\omega'')t} d^3x\,dt \,d^3 k' \,d^3k''\\&= \frac{1}{(2\pi)^3}\int \hat{\rho}_0(\mathbf{k}')\hat{\rho}_1(\omega,\mathbf{k}-\mathbf{k}') d^3 k',
\end{split}
\end{equation}
where we again used the properties of the Dirac delta function in \eqref{newd1}. Then we considered the last term in \eqref{l5} containing $\nabla\Phi_1\cdot\nabla \rho_0$. In this case, we obtained
\begin{equation}
\begin{split}
&\int \nabla\Phi_1\cdot\nabla \rho_0 \,e^{-i(\mathbf{k}\cdot\mathbf{r}-\omega t)} d^3x\, dt=-\frac{1}{(2\pi)^7} \times \\& \int \mathbf{k}''\cdot\mathbf{k}'\hat{\rho}_0(\mathbf{k}')\hat{\Phi}_1(\omega'',\mathbf{k}'') e^{i(\mathbf{k}''+\mathbf{k'}-\mathbf{k})\cdot \mathbf{r}} e^{i(\omega-\omega'')t} d^3x\,dt \,d^3 k' \,d^3k''\\&= \frac{4\pi G}{(2\pi)^3}\int \frac{\mathbf{k}'\cdot(\mathbf{k}-\mathbf{k'})}{(\mathbf{k}-\mathbf{k'})^2} \hat{\rho}_0(\mathbf{k}')\hat{\rho}_1(\omega,\mathbf{k}-\mathbf{k}') d^3 k',
\end{split}
\end{equation}
where we used \eqref{l6}. Finally, using equations \eqref{j13} and \eqref{l6}, and noting that hydrostatic equilibrium implies $\nabla p_0 = -\rho_0 \nabla \Phi_0$, Applying the same Fourier transform procedure to the remaining terms in Eq.~\eqref{l5} yields:
\begin{equation}\label{j14}
\omega^2=\frac{1}{(2\pi)^3}\frac{1}{\hat{\rho}_1(\omega,\mathbf{k})}\int \mathcal{A}(\mathbf{k},\mathbf{k'})\,\hat{\rho}_1(\omega,\mathbf{k}-\mathbf{k'})\, d^3k' ,
\end{equation}
where the function $\mathcal{A}(\mathbf{k},\mathbf{k'})$ is given by
\begin{equation}\label{j15}
 \mathcal{A}(\mathbf{k},\mathbf{k'})=k^2 \hat{c_s^2}(\mathbf{k'})-4\pi G\hat{\rho}_0(\mathbf{k'})\Big[\frac{\mathbf{k}\cdot\mathbf{k'}}{k'^2}+\frac{\mathbf{k}\cdot(\mathbf{k}-\mathbf{k'})}{(\mathbf{k}-\mathbf{k'})^2}\Big].
\end{equation}
The complex nature of the dispersion relation \eqref{j14} arises because the background functions, including the density and its derivatives, are treated as functions of $\mathbf{r}$. Consequently, it is not possible to derive a general stability condition valid for arbitrary hydrostatic background equilibria from this dispersion relation. For separable perturbations of the form $\rho_1(\mathbf{r},t)=f(\mathbf{r})g(t)$, the situation simplifies considerably. In this case, $\omega$ does not appear on the RHS of equation \eqref{j14}. Consequently, the stability condition $\omega^2 \geq 0$ takes the following form:
\begin{equation}
\frac{1}{\hat{\rho}_1(\mathbf{k})}\int \mathcal{A}(\mathbf{k},\mathbf{k'})\hat{\rho}_1(\mathbf{k}-\mathbf{k'}) d^3k'\geq 0.
\end{equation}
It remains challenging to derive a clear stability bound on $k$ from this simplified form. 
\subsection{An averaging procedure to include background gradients}
Fortunately, an approximate scheme for local perturbations exists that circumvents these complexities and yields a simpler stability criterion. The principal equation \eqref{l5} constitutes an integro-differential equation for the perturbation $\rho_1$. Note that $\Phi_1$ is related to $\rho_1$ via $\Phi_1\propto \int k^{-2}\hat{\rho}_1(\omega,\mathbf{k}) e^{i(\mathbf{k}\cdot\mathbf{r}-\omega t)}d^3 k d\omega$. All coefficients appearing in the integro-differential equation for $\rho_1$ are functions of $\mathbf{r}$. As previously mentioned, this is the origin of the complexity. We employ an averaging procedure to simplify the problem. Specifically, after expressing $\rho_1$ in Fourier integral form, we first multiplied equation \eqref{l5} by $e^{i\omega t}$ and integrated over time. We then computed the spatial average of all terms, denoting these spatial averages as $\langle Q(\mathbf{r}) \rangle = \frac{1}{\delta V} \int Q(\mathbf{r'}) d^3 x'$, where $\delta V$ is a small characteristic volume $4\pi a^3/3$ enclosing the perturbation. We assumed that $\mathbf{r}_0$ specified the location of the perturbation and that the averaging is performed around this point. For example, for the first term in \eqref{l5}, we obtained
\begin{equation}
\langle \int \frac{\partial^2\rho_1}{\partial t^2}e^{i \omega t} dt\rangle=-\frac{\omega^2}{(2\pi)^3\delta V}\int \hat{\rho}_1\int_{\delta V}e^{i\mathbf{k}\cdot\mathbf{r'}}d^3 x' d^3 k,
\end{equation}
where we wrote $\mathbf{r'}=\mathbf{r}_0+\mathbf{x}$ with $\mathbf{x}$ varying inside the sphere of radius $a$. Note that $\hat{\rho}_1$ is a function of $\omega$ and $\mathbf{k}$. We omitted the explicit arguments $\hat{\rho}_1(\omega,\mathbf{k})$ for notational brevity. The result takes the form:
\begin{equation}\label{n1}
\langle \int \frac{\partial^2\rho_1}{\partial t^2}e^{i \omega t} dt\rangle\propto-\omega^2\int  \hat{\rho}_1 e^{i \mathbf{k}\cdot\mathbf{r}_0} I_1(ka) d^3k.
\end{equation}
We used the symbol $\propto$ because the coefficient $\frac{1}{(2\pi)^3\delta V}$ was omitted. Furthermore, $I_1(ka)$ is defined as:
\begin{equation}
I_1(ka)=\int_{\delta V} e^{i \mathbf{k}\cdot \mathbf{x}} d^3x=\frac{4\pi}{k^3}(\sin ka- ka \cos ka).
\end{equation}
This integral is evaluated in spherical coordinates by aligning $\mathbf{k}$ along the $z$ direction. As another example, we considered the third term involving $\rho_0\rho_1$:
\begin{equation}
\langle\int \rho_0\rho_1e^{i\omega t}dt\rangle\propto \int \hat{\rho}_1\int_{\delta V}\rho_0(\mathbf{r'})e^{i\mathbf{k}\cdot\mathbf{r'}}d^3 x' d^3k.
\end{equation}
We then employed the Taylor expansion of the density, $\rho_0(\mathbf{r'})\simeq \rho_0(\mathbf{r}_0)+\nabla\rho_0(\mathbf{r}_0)\cdot\mathbf{x}+\mathcal{O}(|\mathbf{x}|^2)$, which is appropriate for local perturbations. The result takes the form:
\begin{equation}\label{new1}
\begin{split}
\langle\int \rho_0\rho_1e^{i\omega t}dt&\rangle\propto \int \hat{\rho}_1 e^{i\mathbf{k}\cdot\mathbf{r}_0} \rho_{0}(\mathbf{r_0})I_1(ka)d^3k\\
&+i \int \hat{\rho}_1 e^{i\mathbf{k}\cdot\mathbf{r}_0} \mathbf{k}\cdot\nabla\rho_{0}(\mathbf{r_0})I_2(ka)d^3k,
\end{split}
\end{equation}
where $I_2(ka)$ is defined as follows:
\begin{equation}
\int_{\delta V}\mathbf{x}e^{i\mathbf{k}\cdot\mathbf{x}}d^3x=i \mathbf{k} I_2(ka).
\end{equation}
we can simplify equation ~\eqref{new1} Similar to $I_1(ka)$, this integral is simply evaluated in spherical coordinates by aligning $\mathbf{k}$ along the $z$ direction. The result is
\begin{equation}
I_2(ka) =-\frac{4\pi}{k^5}\Big[3ka \cos ka+(k^2a^2-3)\sin ka\Big].
\end{equation}
we can simplify equation \eqref{new1} as follows
\begin{equation}\label{n2}
\begin{split}
\langle\int \rho_0\rho_1e^{i\omega t}dt&\rangle\propto\int \hat{\rho}_1 e^{i\mathbf{k}\cdot\mathbf{r}_0} \rho_{0}(\mathbf{r_0})I_1(ka)[1+i \alpha_1 \mathcal{I}]d^3k,
\end{split}
\end{equation}
where $\alpha_1(\mathbf{k})=\frac{\mathbf{k}\cdot\nabla \rho_0}{k^2\rho_0}$ and $\mathcal{I}(ka)=k^2 I_2(ka)/I_1(ka)$ and can be written as:
\begin{equation}
\mathcal{I}(ka)=3+\frac{k^2a^2}{ka \cot ka-1}.
\end{equation}
We next considered the term $c_s^2 \nabla^2 \rho_1$. In this case, we obtained
\begin{equation}\label{n3}
\langle \int c_s^2 \nabla^2 \rho_1 e^{i\omega t}dt\rangle\propto -\int \hat{\rho}_1 e^{i\mathbf{k}\cdot\mathbf{r}_0} I_1 k^2 c_s^2(\mathbf{r}_0) [1+i \alpha_2 \mathcal{I}] d^3k,
\end{equation}
where $\alpha_2(\mathbf{k})=\frac{\mathbf{k}\cdot\nabla c_s^2(\mathbf{r}_0)}{k^2 c_s^2(\mathbf{r}_0)}$. Similarly for the terms $\rho_1 \nabla^2 c_s^2$ and $\nabla c_s^2\cdot \nabla \rho_1$ we obtained:
\begin{equation}
\begin{split}\label{new3}
&\langle \int \rho_1\nabla^2 c_s^2 e^{i\omega t}dt\rangle\propto \int \hat{\rho}_1 e^{i\mathbf{k}\cdot\mathbf{r}_0} I_1 \nabla^2c_s^2(\mathbf{r}_0) [1+i \alpha_3 \mathcal{I}] d^3k,\\&
\langle \int \nabla c_s^2\cdot \nabla \rho_1 e^{i\omega t}dt\rangle\propto \int \hat{\rho}_1 e^{i\mathbf{k}\cdot\mathbf{r}_0} I_1 k^2 c_s^2(\mathbf{r}_0) [i \alpha_2-\beta_2] d^3k,
\end{split}
\end{equation}
where $\alpha_3(\mathbf{k})=\frac{\mathbf{k}\cdot\nabla [\nabla^2 c_s^2(\mathbf{r}_0)]}{k^2 \nabla^2 c_s^2(\mathbf{r}_0)}$ and $\beta_2$ is defined as 
\begin{equation}
\beta_2(\mathbf{k})=\sum_{j=1}^{3}\frac{k_j}{k}\frac{\mathbf{k}\cdot\nabla (\partial c_s^2/\partial x^j)}{k^3 c_s^2}|_{\mathbf{r}_0},
\end{equation}
and we used $\int_{\delta V}x_j  \,e^{i\mathbf{k}\cdot\mathbf{x}}d^3x=i k_j\, I_2(ka)$ for the derivation of the second relation in equation \eqref{new3}. So far we have considered contributions from the first three terms in equation \eqref{l5}. We now consider the final two terms. Using the divergence theorem, we found that the fourth term in \eqref{l5} vanishes upon integration over $\delta V$. Here, $\delta V$ denotes a volume encompassing the perturbation, and we therefore expect $\rho_1$ to vanish on the boundary of this volume. On the other hand, for the last term we obtained
\begin{equation}\label{n4}
\langle \int \nabla\Phi_1\cdot\nabla \rho_0 e^{i\omega t}dt\rangle\propto 4\pi G\int \hat{\rho}_1 e^{i\mathbf{k}\cdot\mathbf{r}_0} I_1 \rho_0(\mathbf{r}_0) [\beta_1-i\alpha_1] d^3k,
\end{equation}
where $\beta_1$ is defined as
\begin{equation}
\beta_1(\mathbf{k})=\sum_{j=1}^{3}\frac{k_j}{k}\frac{\mathbf{k}\cdot\nabla (\partial \rho_0/\partial x^j)}{k^3 \rho_0}|_{\mathbf{r}_0}.
\end{equation}
Now, combining equations \eqref{n1}, \eqref{n2}, \eqref{n3}, \eqref{new3}, and \eqref{n4} into equation \eqref{l5}, and after straightforward algebra, we obtain an integral over $\mathbf{k}$ as follows:
\begin{equation}
\int \hat{\rho}_1(\omega,\mathbf{k})I_1(ka)e^{i\mathbf{k}\cdot\mathbf{r}_0}(\omega^2-\mathcal{B}(\mathbf{k}))d^3k =0.
\end{equation}
This holds for any local perturbation $\hat{\rho}_1(\omega,\mathbf{k})$ satisfying our approximation. Setting the integrand to zero yields the following dispersion relation $\omega^2=\mathcal{B}(\mathbf{k})$:
\begin{equation}
\begin{split}\label{lp6}
\omega^2(\mathbf{k})=c_s^2 k^2\Big[1+&i\,(\mathcal{I}-2)\alpha_2(\mathbf{k})+2\beta_2(\mathbf{k})\mathcal{I}\Big]\\&-4\pi G\rho_0 \Big[1+i\,(\mathcal{I}-1)\alpha_1(\mathbf{k})+\beta_1(\mathbf{k}) \mathcal{I}\Big]\\&
-\nabla^2c_s^2 \Big[1+i \mathcal{I}\alpha_3(\mathbf{k})\Big].
\end{split}
\end{equation}
This is the main dispersion relation derived in this paper. Since $\alpha_i$, $\beta_i$, and $\mathcal{I}$ are functions of $\mathbf{k}$, the dispersion relation \eqref{lp6} is highly nontrivial. It should also be noted that the characteristic size of the perturbation, namely $a$, also appears in the dispersion relation through the function $\mathcal{I}(k a)$. In the limit where the spatial gradients of the background quantities are neglected, this dispersion relation reduces to the standard case. 

\subsection{Stability of local short-wavelength perturbations}
We now express the previously defined functions
$\alpha_i(\mathbf{k})$ and $\beta_i(\mathbf{k})$
in the following compact form:
\begin{equation}
\begin{split}
&\alpha_i(\mathbf{k})=\frac{\mathbf{k}\cdot\nabla \psi_i}{k^2\psi_i},\\&
\beta_i(\mathbf{k})=\sum_{j=1}^{3}\frac{k_j}{k}\frac{\mathbf{k}\cdot\nabla (\partial\psi_i/\partial x^j)}{k^3\psi_i},
\end{split}
\end{equation}
where $\psi_i\equiv(\rho_0,c_s^2,\nabla^2c_s^2)$. Of course $\beta_3$ does not appear in our calculations. All these functions in the dispersion relation are evaluated at $\mathbf{r}_0$. The functions $\alpha_i(\mathbf{k})$ can be approximated as
\begin{equation}\label{new56}
\alpha_i(\mathbf{k})= \frac{\mathbf{k}\cdot\mathbf{n}}{k^2}\frac{|\nabla\psi_i|}{\psi_i}\simeq\frac{\cos\theta}{k l},
\end{equation}
where the unit vector $\mathbf{n}$ indicates the direction of the $\psi_i$ gradient. On the other hand, we approximated $\frac{|\nabla\psi_i|}{\psi_i}\approx \frac{1}{l}$, where $l$ represents the characteristic length scale over which background quantities such as the density $\rho_0$ vary in the unperturbed configuration at the location of the perturbation. Similarly, for the functions $\beta_i(\mathbf{k})$ we wrote
\begin{equation}\label{lpp6}
\beta_i(\mathbf{k})\simeq\sum_{j=1}^{3}\frac{k_j}{k}\frac{\cos\theta_j}{k^2 l^2},
\end{equation}
where we assumed $|\nabla(\partial\psi_i/\partial x^j)|\approx\psi_i/l^2$, and $\theta_j$ is the angle between $\mathbf{k}$ and $\nabla(\partial\psi_i/\partial x^j)$. The evolution of perturbations with wavelengths much shorter than the system size is of particular interest. In other words, we focused on the dynamics of local perturbations rather than the global stability of the host system. Accordingly, we assumed $kl\gg 1$ (consistent with our averaging procedure) and neglected all the $\alpha_i$ and $\beta_i$ terms in the dispersion relation, as they scale as $1/kl$ and $1/k^2l^2$, respectively. Note that $|k_j/k|<1$ and also $\mathcal{I}(k a)<1$ for $k \simeq 1/a$, which is the scale of the local perturbation. Thus, in this limit, the dispersion relation reduces to
\begin{equation}
\omega^2\simeq c_s^2 k^2-4\pi G\rho_0 -\nabla^2 c_s^2.
\end{equation}
However, the last term on the right-hand side is much smaller than the first term. To see this explicitly, their ratio is $\frac{\nabla^2 c_s^2}{c_s^2 k^2}\approx \frac{1}{k^2 l^2}$. Therefore, in the short-wavelength regime, we therefore also neglected the last term and finally recover the following dispersion relation $\omega^2\simeq c_s^2 k^2-4\pi G\rho_0$. This is the standard dispersion relation in the Jeans analysis but obtained without Jeans swindle. Therefore, we conclude that for local short-wavelength perturbations, the nonzero spatial gradients of the background quantities do not change the stability criterion. The standard Jeans analysis therefore remains valid. This is in agreement with the results of \cite{nipoti} where the same topic, namely Jeans analysis including density and pressure gradients, has been investigated. We emphasize that our stability analysis so far is restricted to short-wavelength perturbations $k l\gg 1$ or equivalently $\lambda\ll l$. In this case if $\lambda<\lambda_{\rm J}$ then the perturbation is stable and for the case $\lambda>\lambda_{\rm J}$ we have instability. The latter case may occur in systems where the Jeans wavelength is much smaller than the characteristic size of the system.  

Keeping terms up to $1/kl$ and neglecting higher-order terms of $\mathcal{O}(1/k^2l^2)$ and $\mathcal{O}(1/k^3l^3)$, the dispersion relation \eqref{lp6} reduces to
\begin{equation}
\begin{split}
\omega^2(\mathbf{k}) \simeq c_s^2 k^2 - 4\pi G \rho_0 + & i\, \Big[ c_s^2 k^2(\mathcal{I}-2)\alpha_2(\mathbf{k})\\& - 4\pi G \rho_0 (\mathcal{I}-1)\alpha_1(\mathbf{k})\Big],
\end{split}
\end{equation}
which is still fairly complicated. Using Eq.~\eqref{new56}, the dispersion relation above can be approximated as
\begin{equation}
\omega^2(\mathbf{k}) \simeq c_s^2 k^2 - 4\pi G \rho_0 +  i\, \frac{\cos\theta}{k l}\Big[ c_s^2 k^2(\mathcal{I}-2) - 4\pi G \rho_0 (\mathcal{I}-1)\Big],
\end{equation}
where $\theta$ is the angle between $\mathbf{k}$ and $\nabla\rho_0$. It should be noted that $\nabla c_s^2 = \frac{dc_s^2}{d\rho_0}\nabla\rho_0 = \frac{d^2 p_0}{d\rho_0^2}\nabla\rho_0$. As a further valid approximation, since $ka\simeq 1$, we adopted an average value $\mathcal{I}\simeq 1$, which simplified the dispersion relation to
\begin{equation}
\omega^2(\mathbf{k}) \simeq c_s^2 k^2 - 4\pi G \rho_0 - i\,c_s^2 \cos\theta \frac{k}{l}.
\end{equation}
If the self-gravity of the background is neglected, this dispersion relation becomes identical to that of sound waves in a stratified medium (see Chapter 6 of \citealt{2007pafd.book.....C}). We then considered the special case of wavenumbers $\mathbf{k}$ perpendicular to the gradient of the background density at the point $\mathbf{r}_0$, namely $\mathbf{k}\perp\nabla\rho_0$. Interestingly, for these perturbation modes the standard dispersion relation, and consequently the Jeans criterion, remain valid. For the case $\cos\theta\neq 0$, however, assuming that $k$ is real, it is clear that $\omega$ has both real and imaginary parts, i.e., $\omega=\omega_{\rm R}+i\omega_{\rm I}$. It is straightforward to show that $\omega_{\rm I}$ can be either positive or negative: negative values correspond to damped oscillatory behavior, whereas $\omega_{\rm I}>0$ yields growing oscillatory solutions. Solutions of this type are referred to as overstable.

Conversely, for real $\omega$, the wavenumber $k$ has an imaginary part given by:
\begin{equation}
k(\omega)=\pm \sqrt{\Big(k_{\rm J}^2+\frac{\omega^2}{c_s^2}\Big)-\frac{\cos^2\theta}{4 l^2}}+i\frac{\cos\theta}{2l}.
\end{equation}
Assuming that the argument of the square root is positive, we can write $k(\omega)=k_{\rm R}(\omega)+i\, k_{\rm I}$. In this case, the corresponding perturbation mode $\rho_1$ can be expressed as
\begin{equation}
\rho_1\propto e^{-k_{\rm I}(\mathbf{n}\cdot\mathbf{r})} e^{i[k_{\rm R}(\omega)(\mathbf{n}\cdot\mathbf{r})-\omega t]},
\end{equation}
where $\mathbf{n}$ is the unit vector along the direction of $\nabla\rho_0$. When $\Big(k_{\rm J}^2+\frac{\omega^2}{c_s^2}\Big)-\frac{\cos^2\theta}{4 l^2}<0$, the wavenumber becomes purely imaginary. Writing it as $k=i k_{\rm I}$, where $k_{\rm I}$ is distinct from the definition above, the perturbation mode then takes the form $\rho_1\propto e^{-k_{\rm I} (\mathbf{n}\cdot\mathbf{r})} e^{-i\omega t}$. This does not represent a propagating wave but rather a standing wave. In summary, the stability of an arbitrary perturbation composed of different modes is governed primarily by the wave vectors perpendicular to the gradient of the background density, while other wave vectors give rise to overstable or standing-wave behavior. Thus, notably, even in this limit where terms of order $1/kl$ are retained, the standard Jeans criterion remains valid.

For local perturbations, not only should $k$ be larger than $1/l$, but it should also be larger than $1/\mathcal{L}$, where $\mathcal{L}$ is the characteristic length scale of the background system. Using the virial theorem, one can show that $1/\mathcal{L}\sim k_{\rm J}$ \citep{binney,2023MNRAS.518.5154N}. Therefore, for local perturbations we have $k>k_{\rm J}$. On the other hand, as clarified above, even in the presence of $1/kl$ terms in the dispersion relation, the standard Jeans criterion remains valid; in other words, waves with $k>k_{\rm J}$ are stable. Since all local perturbations have wavenumbers larger than the Jeans wavenumber, it follows that all local perturbations in a hydrostatic stratified system are stable. This result agrees with that of \citet{nipoti}, who employs a different approach to investigate gravitational stability in a stratified medium.

As a final remark in this section, the question remains: what happens to long-wavelength perturbations? When $kl\ll 1$, our analysis no longer applies. Consequently, we cannot draw definitive conclusions regarding the stability of long-wavelength modes.

\section{Conclusions}\label{section4}
For the sake of completeness, we first employed physical arguments to demonstrate that the standard Jeans analysis remains valid for linear perturbations whose characteristic size is small compared to the characteristic scale of the host system, such that the gravitational field of the background does not affect the internal dynamics of the perturbation. In this regime, the spatial gradients of the perturbed quantities are expected to substantially exceed those of the background, so that no Jeans swindle is required to derive the Jeans criterion. As the main objective of this paper, we then relaxed these assumptions and incorporated the background density and pressure gradients, as well as the background gravitational field. We implemented an averaging procedure to incorporate the background gradients, from which we derived a new dispersion relation given by equation \eqref{lp6}. This dispersion relation is considerably complicated, in the sense that the direction of the wavevector matters, not merely its magnitude. Both the gradients of the background quantities and the characteristic size of the perturbation itself appear explicitly in the dispersion relation, making it impossible to derive a general stability criterion valid for an arbitrary hydrostatic system.

However, by expanding the dispersion relation in terms of $\epsilon=1/kl$ and retaining terms up to first order in $\epsilon$, we showed that for short-wavelength local perturbations, the dispersion relation \eqref{lp6} reduces to the standard Jeans criterion. We also showed that, in agreement with \cite{nipoti}, all local perturbations in a stratified hydrostatic medium are stable against gravitational collapse. It is worth reiterating that, even when terms of order $1/kl$ are retained in the dispersion relation, the standard Jeans analysis remains valid. Our approach, however, is not applicable to long-wavelength perturbations.

\begin{acknowledgements}
We thank the referee, Alessandro Romeo, for his valuable comments, which helped us substantially improve the manuscript. We are grateful to Prof. Bahram Mashhoon for his valuable comments. We also thank Tahere Kashfi for reading an early version of the draft and providing helpful comments. This work is supported by Ferdowsi University of Mashhad.
\end{acknowledgements}
\bibliographystyle{aa}
\bibliography{short,Jeans}

\end{document}